\documentclass[12pt]{article}
\usepackage{amsmath}
\usepackage{amsfonts}
\usepackage{graphicx,psfrag,epsf}
\usepackage{enumerate}
\usepackage{natbib}
\usepackage{url} 
\usepackage{bm}

\newcommand{\blind}{0}

\begin{document}

\def\spacingset#1{\renewcommand{\baselinestretch}%
{#1}\small\normalsize} \spacingset{1}





\if0\blind
{
  \title{\bf Approximate Bayesian Computation and Model Assessment for Repulsive Spatial Point Processes}
  \author{Shinichiro Shirota\thanks{PhD student, Department of Statistical Science, Duke University, U.S. (E-mail: ss571@stat.duke.edu)}\hspace{.2cm}\\
    Department of Statistical Science, Duke University, U.S. \\
    and \\
    Alan. E. Gelfand\thanks{Professor, Department of Statistical Science, Duke University, U.S. (E-mail: alan@stat.duke.edu)} \\
    Department of Statistical Science, Duke University, U.S.}
  \maketitle
} \fi

\if1\blind
{
  \bigskip
  \bigskip
  \bigskip
  \begin{center}
    {\LARGE\bf Approximate Bayesian Computation and Model Assessment for Repulsive Spatial Point Processes}
\end{center}
  \medskip
} \fi

\bigskip
\begin{abstract}
In many applications involving spatial point patterns, we find evidence of inhibition or repulsion.  The most commonly used class of models for such settings are the Gibbs point processes.  A recent alternative, at least to the statistical community, is the determinantal point process.  Here, we examine model fitting and inference for both of these classes of processes in a Bayesian framework.  While usual MCMC model fitting can be available, the algorithms are complex and are not always well behaved.  We propose using approximate Bayesian computation (ABC) for such fitting. This approach becomes attractive because, though likelihoods are very challenging to work with for these processes, generation of realizations given parameter values is relatively straightforward.  As a result, the ABC fitting approach is well-suited for these models.  In addition, such simulation makes them well-suited for posterior predictive inference as well as for model assessment.  We provide details for all of the above along with some simulation investigation and an illustrative analysis of a point pattern of tree data exhibiting repulsion. R-code and datasets are included in the supplementary material.
\end{abstract}

\noindent%
{\it Keywords:}  determinantal point processes; Gibbs point processes; model checking; summary statistics
\vfill

\newpage
\spacingset{1.45} 

\section{Introduction}
\label{sec:intro}
There is increasing interest in analyzing spatial point process data.  In the literature, the most widely adopted class of models are nonhomogeneous Poisson processes (NHPP) or, more generally, Cox processes, including log Gaussian Cox processes (LGCP) (see \cite{MollerWaagepetersen(04)} and references therein).  Such models assume conditionally independent event locations given the process intensity.  However, in many applications, we find evidence of clustering or of inhibition.  Here, we focus on inhibition and refer to associated models as repulsive spatial point processes.   Most common in this setting are the Gibbs point processes (here, denoted as GPP) (see, e.g., \cite{Illianetal(08)}). These processes specify the joint location density, up to normalizing constant, in the form of a Gibbs distribution, introducing potentials on cliques of order $1$ but also potentials on cliques of higher order, which capture interaction.  The most familiar example in the literature is the Strauss process and its extreme version, the hardcore process  (\cite{Strauss(75)} and \cite{KellyRipley(76)}).  An attractive alternative is the determinantal point process (here, denoted as DPP).  Though these processes have some history in the mathematics and physics communities, they have only recently come to the attention of the statistical community thanks, most notably, to recent efforts of Jesper M{\o}ller and colleagues.  See, for instance, \cite{Lavancieretal(15)}.

The contribution of this paper is to explore both the GPP and DPP models with regard to application.  We will briefly summarize and compare their respective properties.  We will consider model fitting in a Bayesian framework.  Markov chain Monte Carlo (MCMC) model fitting has been proposed for both GPPs and DPPs (\cite{Molleretal(06)}, \cite{Affandietal(13)}, \cite{Affandietal(14)}, \cite{Goldsteinetal(15)}).  The algorithms are quite complex and implementation can often result in poorly behaved chains with concerns regarding posterior convergence. Here, we propose much simpler model fitting using approximate Bayesian computation (ABC).  ABC is particularly promising for GPPs and DPPs since these processes allow straightforward simulation of point pattern realizations given parameter values.  Additionally, such simulation facilitates posterior inference as well as consideration of model adequacy and model comparison.

ABC methods are now attracting considerable attention (\cite{Pritchardetal(99)}, \cite{Beaumontetal(02)}, \cite{Marjorametal(03)}, \cite{SissonFan(11)} and \cite{Marinetal(12)}).
The scope of ABC applications is also increasing, e.g., population genetics (\cite{Beaumontetal(02)}), multidimensional stochastic differential equations (\cite{Picchini(13)}), macroparasite populations (\cite{DrovandiPettitt(11)}) and the evolution of the HIV-AIDS (\cite{BlumTran(10)}).
As for spatial statistical application, \cite{ErhardtSmith(12)} implemented ABC for max-stable processes in order to model spatial extremes and \cite{Soubeyrandetal(13)} applied ABC with functional summary statistics to fit a cluster and a marked spatial point process.

We briefly review some of the repulsive point process literature. The Gibbs point process offers a mechanistic model with interpretable parameters and has been used for modeling repulsive point patterns in environmental science and biology (\cite{StoyanPenttinen(00)}, \cite{Mattfeldtetal(07)}, \cite{Kingetal(12)} and \cite{Goldsteinetal(15)}).
The main challenge for fitting models using these processes is that likelihoods have intractable normalizing constants which depend on parameters. Hence, likelihood inference is difficult (\cite{MollerWaagepetersen(04)}) and standard Bayesian inference using  Markov chain Monte Carlo (MCMC)) is not directly available.
Maximum pseudo-likelihood estimation was proposed in \cite{Besag(75)}, \cite{Besag(86)} and \cite{JensenMoller(91)}. These estimators show poor performance in the presence of strong inhibition (e.g., \cite{HuangOgata(99)}).
In the Bayesian framework, a clever auxiliary variable MCMC strategy has been developed by \cite{Molleretal(06)} and extended by \cite{Murrayetal(06)}. However, perfect simulation within the MCMC algorithm is needed along with approximations.

Determinantal point processes arise in random matrix theory and quantum physics, and are investigated mainly by probabilists with a machine learning perspective (see, \cite{Macchi(75)}, \cite{Houghetal(06)} and \cite{KuleszaTasker(12)}).
In contrast to GPPs, DPPs have analytical expressions for moments and approximate likelihoods. In particular, \cite{Lavancieretal(15)} investigated statistical inference for and properties of spatial DPPs along with approximation for evaluation of the likelihood.
The pattern of repulsiveness exhibited by DPPs may be different from that of GPPs so, making a modeling choice for a spatial repulsive point process may be unclear.

We propose to implement the ABC algorithm based on ABC-MCMC proposed by \cite{Marjorametal(03)} for fitting both GPPs and DPPs. We include discussion about how to choose summary statistics, kernel functions and, tuning user specific parameters.
Again, the attractiveness of ABC for repulsive point processes rests in the fact that it is straightforward to generate realizations under these point processes given parameter values.
This enables the ABC presumption: randomly draw parameters and then randomly draw point patterns given parameters.
Further, with posterior inference achieved for the model parameters, we can use composition sampling to draw posterior predictive point patterns enabling posterior inference about features of point patterns realized under the models. In addition, through these posterior samples of point patterns, we can propose model assessment for repulsive point processes, following the discussion in \cite{LeiningerGelfand(16)}.

We offer simulation investigation of the model fitting approach. We also analyze a real dataset consisting of 89 trees from the Blackwood region in Duke Forest in the US.
All of the statistical analyses presented here were implemented with R. In particular, the $\texttt{spatstat}$ package (\cite{BaddeleyTurner(05)}) is utilized for simulating point patterns and calculating functional summary statistics associated with DPPs and GPPs. The R code is attached as supplementary material.

The format of the paper is as follows.  Section \ref{sec:rspp} introduces the repulsive point processes we investigate and their associated inference. Section \ref{sec:abcrpp} provides the development of our ABC fitting strategy for repulsive point processes along with model validation for and comparison strategy between different repulsive point processes. In Section \ref{sec:simulation}, we investigate the proposed algorithms with some simulation studies while in Section \ref{sec:real} we analyze the foregoing forest data. Finally, Section \ref{sec:disc} offers discussion about the pros and cons of our approach as well as potential future work.
\section{Repulsive Spatial Point Processes}
\label{sec:rspp}

Here, we provide a brief review of the two classes of repulsive point processes that we consider.  GPPs are taken up in Section \ref{sec:gpp}, DPPs in Section 2.2.
We consider a bounded spatial domain $\it{D}$ $\subset$ $\mathbb{R}^{2}$ and denote the (finite) point pattern over $\it{D}$ by $\bm{x}=\{x_{1}, \ldots, x_{n}\}$.
\subsection{Gibbs Point Processes}
\label{sec:gpp}
The joint location density for a Gibbs process takes the form of a Gibbs distribution (e.g., \cite{Georgii(88)}).  In particular, we say that a point process model is a finite
Gibbs process if, for $n$ locations, its location density is
\begin{equation}
\pi(\bm{x})= \mbox{exp}(-Q(\bm{x}))
\end{equation}
with regard to a homogeneous Poisson process (HPP) with unit intensity (e.g., \cite{vanLieshout(00)} or \cite{MollerWaagepetersen(04)}).   In general,
\begin{equation}\label{GibbsP2}
Q(x_{1},x_{2},...,x_{n})= c_{0} + \sum_{i=1}^{n} h_{1}(x_{i}) + \sum_{i \neq j} h_{2}(x_{i}, x_{j}) + ... + h_{n}(x_{1},x_{2},...,x_{n}).
\end{equation}
In (\ref{GibbsP2}), the $h$'s are potentials of order $1$, $2$,...$n$, respectively, each symmetric in its arguments.  Here, $c_{0}$ is a \emph{normalizing} constant to make $\pi(\bm{x})$ integrate to 1 over $D^{n}$.  Denoting parameters in the Gibbs potentials by $\bm{\theta}$, the normalizing constant becomes $c_{0}(\bm{\theta})$. Evidently it can not be calculated analytically and, in fact, is computationally intractable.

With potentials only of order $1$, we obtain a nonhomogeneous Poisson process (NHPP) with $\lambda(x) \propto e^{-h_{1}(x)}$.  Higher order potentials capture/control interaction; customarily only $h_{1}$ and $h_{2}$ are included.  To guarantee integrability, we must take $h_{2} \geq 0$ which implies that we can only capture inhibition.  In other words, if we require $Q(x_{1},x_{2}) \geq c_{0} + h_{1}(x_{1}) + h_{1}(x_{2})$, this means for pairs of points at a given distance, $\pi(x_{1},x_{2})$ puts less mass under the Gibbs specification than with the corresponding NHPP; we encourage inhibition.  If $h_{1}(x)$ is constant, we have a homogeneous Gibbs process.  The most common form for $h_{2}$ is $\phi(||x - x'||)$, e.g., $\phi(||x - x'||) = e^{-||x -
x'||^{2}/\eta^{2}}$ . The Papangelou conditional intensity (\cite{Illianetal(08)}) becomes
\begin{equation} \label{PapGibbs}
\lambda(x|\bm{x}) = \exp(-(h_{1}(x) + \sum_{i=1}^{n}\phi(\|x - x_{i}\|))).
\end{equation}
Attractively, the unknown normalizing constant cancels in the conditional intensity.

The Strauss process is a GPP with density often written as (e.g., \cite{MollerWaagepetersen(04)})
\begin{align}
\pi(\bm{x})=\beta^{n(\bm{x})}\gamma^{s_{R}(\bm{x})}/c(\beta,\gamma), \quad \bm{x}\in D^{n}
\end{align}
where $\beta>0$, $0\le\gamma \le1$, $n(\bm{x})$ is the number of points, $c(\beta,\gamma)$ is the normalizing constant and
\begin{align}
s_{R}(\bm{x})=\sum_{\{\xi, \eta\}\subseteq \bm{x}}\bm{1}(\|\xi-\eta\|\le R), \quad \xi, \eta \in D
\label{eq:s_R}
\end{align}
is the number of \emph{$R$-close} pairs of points in $\bm{x}$. Given $R$,
$n(\bm{x})$ and $s_{R}(\bm{x})$ are sufficient statistics for $(\beta,\gamma)$. $\gamma$ is an interaction parameter indicating the degree of repulsion.
Large value of $\gamma$ suggest weak repulsion while small values of $\gamma$ indicate strong repulsion. $\gamma=0$ provides the hardcore Strauss process which does not allow  occurrence of any points within radius $R$. $\gamma=1$ provides a homogeneous Poisson process.

For GPPs, since $c_{0}(\bm{\theta})$ cancels out of the Papangelou conditional intensity, the pseudo-likelihood, in log form,  $\log PL(\bm{x}|\bm{\theta})=-\int_{D}\lambda(u|\bm{x}, \bm{\theta})du +\sum_{i} \log \lambda(x_{i}|\bm{x}, \bm{\theta})$, has been proposed (\cite{Besag(75)}) yielding the maximum pseudo-likelihood estimator.
Although the maximum pseudo-likelihood estimator is consistent (see, \cite{JensenMoller(91)} and \cite{Mase(95)}), the performance of the maximum pseudo-likelihood estimator is poor in the case of a small number of points and strong repulsion (\cite{HuangOgata(99)}).

The pseudo-likelihood can be used for MCMC in the Bayesian framework (e.g. \cite{Kingetal(12)}).
\cite{Molleretal(06)} and \cite{BerthelsenMoller(08)} proposed a clever auxiliary variable MCMC method (AVM) where, conveniently, $c_{0}(\bm{\theta})$ cancels out of the Hastings ratio.
The challenge is to obtain the conditional density of the auxiliary variable. A partially ordered Markov model is used to approximate this density. A similar approach is the exchange algorithm proposed by \cite{Murrayetal(06)}. Both algorithms require perfect simulation from the likelihood given $\bm{\theta}$ for each MCMC iteration. Although, perfect simulation is available for GPPs, this step can be computationally burdensome and obtaining a good acceptance rate is difficult.

More recently, \cite{Liang(09)} proposed the double MCMC algorithm which does not require perfect simulation from the likelihood. It only requires simulation from the Markov transition kernel and is faster than the AVM and exchange algorithms but convergence to the stationary distribution is not guaranteed. \cite{Goldsteinetal(15)} implement this algorithm, with an application, for a class of GPP models.

For ABC, we need simulation of realizations of a GPP given parameter values.  This is usually based on a birth-and-death algorithm (e.g., \cite{GeyerMoller(94)}, \cite{MollerWaagepetersen(04)} and \cite{Illianetal(08)}).  The simulation algorithm we use to generate the point pattern is "dominated coupling from the past" (\cite{KendallMoller(00)}) as implemented by \cite{BerthelsenMoller(02)} and \cite{BerthelsenMoller(03)}. This algorithm can be called as a default setting in \texttt{spatstat} (\cite{BaddeleyTurner(05)}).

\subsection{Determinantal Point Processes}
\label{sec:dpp}
Theoretical properties for DPPs are investigated by \cite{Houghetal(06)} and elaborated from a statistical perspective by \cite{Lavancieretal(15)}. The $n$-th order product density function (\cite{MollerWaagepetersen(04)}) for DPPs arises as determinants of covariance kernels. We briefly review the definition and theoretical properties of DPPs drawn from \cite{Lavancieretal(15)}.
\\

Suppose, for a finite spatial point process $\bm{x}$ on a compact set $D \subset \mathbb{R}^{2}$,
\begin{align}
\tau^{(n)}(x_{1},\ldots, x_{n})=\text{det}\{[C](x_{1},\ldots, x_{n}) \}, \quad (x_{1},\ldots, x_{n})\in D^{n}, \quad n=1,2,\ldots
\end{align}
is the $n$-th order product density function. Here, $C(x_{i}, x_{j})$ is a covariance kernel for locations $x_{i}$ and $x_{j}$ and $\text{det}\{[C](x_{1},\ldots, x_{n})\}$ denotes the determinant with $(i,j)$-th entry, $C(x_i, x_j)$.  Then $\bm{x}$ is called a $DPP$ with kernel $C$ restricted to a compact set $D$, and we write $\bm{x}\sim DPP_{D}(C)$. Hence, the first order density function is $\tau(x)=C(x,x)$
and the pair correlation function is
\begin{align}
g(x_{1},x_{2})=1-\frac{C(x_{1},x_{2})C(x_{2},x_{1})}{C(x_{1},x_{1})C(x_{2},x_{2})}, \quad \text{if} \quad C(x_{1},x_{1})>0 \quad \text{and} \quad C(x_{2},x_{2})>0
\end{align}
whereas it is 0 otherwise.
Since $C$ is a covariance kernel, then $\tau^{(n)}(x_{1},\ldots, x_{n})\le \tau(x_{1})\ldots \tau(x_{n})$ for any $n > 1$, implying repulsion, and $g\le 1$ (\cite{Lavancieretal(15)}).

For a given covariance function, existence conditions for the DPP are supplied in
\cite{Lavancieretal(15)}. Here, we confine ourselves to three real valued covariance functions over $d=2$ dimensional space: (1) the Gaussian kernel, $C_{G}(x, y)=\tau \exp(-\|x-y\|^2/\sigma^2)$, (2) the Whittle-Mat\'{e}rn, $C_{WM}(x,y)=\tau \frac{2^{1-\nu}}{\Gamma(\nu)}(\|x-y\|/\sigma)^{\nu} K_{\nu}(\|x-y\|/\sigma)$, $\nu >0$, and (3) the Generalized Cauchy, $C_{C}(x,y)=\tau (1+\|x-y\|^2/\sigma^2 )^{-\nu-1}$, $\nu >0$,
where $\tau$ is the variance parameter. For all three kernels, with $d=2$, the existence of the process is guaranteed when $\sigma\le \sigma_{max}=(U/\tau)^{1/2}$, where $U=\pi^{-1}$ for the Gaussian, $U=\Gamma(\nu)/(4\pi \Gamma(\nu+1))$ for the Whittle-Mat\'{e}rn and $U=\Gamma(\nu+1)/(\pi\Gamma(\nu))= \nu/\pi$ for the Cauchy.

The Gaussian kernel is asserted to provide \emph{moderate} repulsion.  A specification enabling \emph{stronger} repulsion can be obtained through the spectral density.
\cite{Lavancieretal(15)} supply the power exponential DPP model induced by the spectral density,
\begin{align}
\varphi(x)=\tau \frac{\alpha^{2}}{\pi \Gamma(2/\nu+1)}\exp(-\|\alpha x\|^{\nu}), \quad \tau \ge 0, \quad \alpha>0, \quad \nu>0
\end{align}
For fixed $\tau$ and $\nu$, existence is guaranteed when $\alpha\le \alpha_{max} = \sqrt{\Gamma(2/\nu+1)\pi/\tau}$.
$\alpha \approx \alpha_{max}$ shows strong repulsiveness.
In our simulation examples below, we illustrate the use of the DPP with both a Gaussian kernel (DDP-G) and a power exponential (DDP-PE) to enable investigation of both moderate and stronger repulsiveness.
\\

Simulation algorithms for DPPs are proposed in e.g., \cite{Houghetal(06)}, \cite{Scardicchioetal(09)} and \cite{Lavancieretal(15)}.
We employ the algorithm in Theorem 2 of \cite{Lavancieretal(15)} which is based on the spectral decomposition of the associated kernel covariance function.  Recall the spectral representation of the covariance function, $C(x_{i}, x_{j})=\sum_{k=1}^{\infty} \lambda_{k}\psi_{k}(x_i)\bar{\psi}_{k}(x_j)$ where $(x_i, x_j) \in D \times D$. The existence of the DPP is guaranteed if all of the eigenvalues, $\lambda_{k} \leq 1$.
Define the \emph{projection} kernel function as $K(x_{i}, x_{j})=\sum_{k=1}^{\infty}G_{k}\psi_{k}(x_i)\bar{\psi}_{k}(x_j)$.
Here, the $G_{k}$ are independent Bernoulli variables with mean $\lambda_{k}$ for $k=1,2,\ldots$.  The DPP with this projection kernel has the same distribution as the $DPP_{D}(C)$.

Exact simulation of a determinantal point process model involves an infinite series, which has no analytical solution except for a few kernel choices (e.g., \cite{Macchi(75)}), which might be insufficient to describe the interaction structure in real datasets (\cite{Lavancieretal(15)}). 
So, we need a truncation of the infinite sum.
This truncation is implemented by $(E[\tilde{N}(D)]-E[N(D)])<0.01E[N(D)]$ where $N(D)\sim \sum_{k=1}^{\infty}G_{k}$ (so $E[N(D)]=\sum_{k=1}^{\infty}\lambda_{k}$) and the truncation, $\tilde{N}(D)$, is generated by approximate simulation (\cite{Lavancieretal(15)} and \cite{BaddeleyTurner(05)}).
With $\bm{v}(x)=(\psi_{1}(x),\ldots, \psi_{n}(x))^{T}$, the simulation algorithm given by \cite{Lavancieretal(15)} is
\begin{itemize}
\item[1.] Set $n=\sum_{k=1}^{\infty}G_{k}$
\item[2.] Sample $x_{n}$ from the distribution with density $p_{n}(x)=\|\bm{v}(x)\|^2/n$, $x\in D$ and set $\bm{e}_{1}=\bm{v}(x_{n})/\|\bm{v}(x_{n})\|$
\item[3.] For $i=n-1$ to $i=1$, sample $x_{i}$ from the distribution with density
\begin{align*}
p_{i}(x)=\frac{1}{i}\biggl\{\|\bm{v}(x)\|^2-\sum_{j=1}^{n-i}|\bm{e}_{j}^{*}\bm{v}(x)|^2 \biggl\}, \quad x\in D.
\end{align*}
Set $\bm{w}_{i}=\bm{v}(x_{i})-\sum_{j=1}^{n-i}\{\bm{e}_{j}^{*}\bm{v}(x_{i})\}\bm{e}_{j}$, $\bm{e}_{n-i+1}=\bm{w}_{i}/\|\bm{w}_{i}\|$.
\item[4.] Return $\{x_{1},\ldots,x_{n}\}$
\end{itemize}
Rejection sampling is used in Step 3 with a uniform density $\lambda_{0}$ over $D$ and acceptance probability given by $p_{i}(x)/\lambda_{0}$, where $\lambda_{0}$ is an upper bound on $p_{i}(x)$ for $x\in D$.
\\


Turning to inference, continuous DPPs (DPPs defined on continuous space $D\in \mathbb{R}^{d}$), e.g., the spatial DPP of interest to us, typically require the infinite dimensional spectral decomposition, which are not analytically available except for a few kernel choices (e.g., \cite{Macchi(75)}), to evaluate the likelihood and to do exact simulation.
The slice sampling approach by \cite{Affandietal(14)}, calculating upper and lower bounds of DPP likelihoods, can be applied but requires low rank approximation of the spectral representation (\cite{Affandietal(13)}).

For stationary continuous DPPs, the likelihood function for $\bm{x}=\{x_{1},\ldots,x_{n}\} \in D^{n}$ is defined as
\begin{equation}
f(\bm{x})=\exp(|D|-H)\text{det}\{[\tilde{C}](x_{1},\ldots,x_{n})\}
\end{equation}
where $H =-\log[P(N(D)=0)]=-\sum_{k=1}^{\infty}\log(1-\lambda_{k})$ and
$\tilde{C}(x,y)=\sum_{k=1}^{\infty}\tilde{\lambda}_{k}\psi_{k}(x)\bar{\psi}_{k}(y)^{T}$, $x, y \in D$, $\tilde{\lambda}_{k}=\lambda_{k}/(1-\lambda_{k})$.
Since the likelihood involves an infinite dimensional spectral decomposition, \cite{Lavancieretal(15)} consider the maximum likelihood estimator based on an approximate likelihood constructed on a rectangular region by using a Fourier basis (see \cite{Lavancieretal(15)} for details). For a large number of points, calculation of components of the covariance kernel, i.e., $C(x_{i}, x_{j})$ for each pair of $(x_{i},x_{j})$, is computationally costly, even using the Fast Fourier transformation. There is also potential sensitivity to the resolution of the grid.
For parametric families of DPPs, the Papangelou conditional intensity is not easier to calculate than the likelihood itself, so pseudo-likelihood estimates are not easily available as with GPPs. Bayesian inference using MCMC will be even more challenging.

Lastly, as \cite{Lavancieretal(15)} note, handling of the DPP for a non-rectangular region is not clear. Embedding a non-rectangular observation window $W$ in a rectangular region $R$, yielding a missing data problem, is one possible approach.
\section{Approximate Bayesian Computation for Repulsive Point Processes}
\label{sec:abcrpp}
Let $\bm{y}$ be the observed point pattern and $\bm{x}$ be a simulated point pattern.
For a Bayesian model of the form $\pi(\bm{x}|\bm{\theta})\pi(\bm{\theta})$, ABC consists of three steps: (1) generate $\bm{\theta}\sim \pi(\bm{\theta})$, (2) generate $\bm{x}\sim\pi(\bm{x}|\bm{\theta})$, (3) compare summary statistics for the generated $\bm{x}$,  $\bm{T}(\bm{x})$, with those of the observed data, $\bm{T}(\bm{y})$, and  accept $\bm{\theta}$ if $\Psi(\bm{T}(\bm{x}),\bm{T}(\bm{y}))<\epsilon$ for a selected \emph{kernel(distance)} measure $\Psi$.
Accepted $\bm{\theta}$ are samples from the approximate posterior distribution, $\pi_{\epsilon}(\bm{\theta}|\bm{T}(\bm{y}))$. Approximation error relative to the exact posterior distribution $\pi(\bm{\theta}|\bm{y})$ comes from the choice of $\bm{T}(\cdot)$, $\Psi$, and $\epsilon$. If $\bm{T}(\cdot)$ is a sufficient statistic for $\bm{\theta}$, then $\pi(\bm{\theta}|\bm{T}(\bm{y}))=\pi(\bm{\theta}|\bm{y})$ and, given $\Psi$, the only approximation error is from $\epsilon$. Since sufficient statistics are not usually available, the selection of informative summary statistics $\bm{T}(\cdot)$ is critically important. Small values of $\epsilon$ are desired but require more simulation of $\bm{\theta}\sim \pi(\bm{\theta})$ and $\bm{x}\sim\pi(\bm{x}|\bm{\theta})$.
Again, with regard to simulation of $\bm{x}$: (i) for the GPP, we can utilize perfect simulation, (ii) for the DPP, we can use the Fourier basis approximate simulation method by truncating infinite sums for DPPs whose kernels don't have an analytical spectral representation.

\subsection{Summary Statistics}
\label{sec:sstat}

Second order summary statistics play a fundamental role in the statistical analysis of spatial point patterns because they illuminate clustering or inhibition behavior.
The most common choice is  Ripley's $K$-function (\cite{Ripley(76)} and \cite{Ripley(77)}). For a stationary point process, the $K$-function with radius $r$, $K_{r}$, is the \emph{expected} number of the remaining points in the pattern within distance $r$ from a typical point. The empirical estimator of $K_{r}$ is
\begin{align}
\hat{K}_{r}(\bm{x})=|D|\sum_{\xi,\eta\in\mathbf x:\,\xi\not=\eta} \frac{\bm{1}[0<\|\xi-\eta\|\le r]}{n(n-1)}e(\xi, \eta)
\end{align}
where $e(\cdot, \cdot)$ is an edge correction factor (e.g., \cite{Illianetal(08)}).
The variance stabilized version,
the $L$-function (\cite{Besag(77)}), $\hat{L}_{r}(\bm{x})=\sqrt{\hat{K}_{r}(\bm{x})/\pi}$ is often preferred. 
Under the Poisson process model, the expected value of $\hat{L}_{r}(\bm{x})$ is $r$.
When $\hat{L}_{r}(\bm{x})-r <0$ for small to modest values of $r>0$, the point pattern suggests an inhibitive point process model (see e.g. \cite{MollerWaagepetersen(04)}).

With the Strauss GPP, given the interaction radius $R$, from (\ref{eq:s_R}), $s_{R}(\bm{x})$, is a sufficient statistic and hence, the appropriate summary statistics would be $\bm{T}=(\log n(\bm{x}), K_{R}(\bm{x}))$.
In practice, $R$ is not known but we can choose a radius $R$ though profile pseudo likelihoods. Alternatively, creating a set of $R$ values yields a set of summary statistics.

For DPPs, repulsiveness of DPPs is determined by the covariance kernel function; there is no notion of a radius. Nonetheless, we propose to use a set $\{\hat{K}_{r}(\bm{x})\}$  evaluated at $M$ selected values of $r$ over the range $[0, r_{max}]$.  That is, with $r_{0}=0, r_{1}, \ldots r_{M}=r_{max}$, calculate $\hat{K}_{r_{1}}(\bm{x}),\ldots, \hat{K}_{r_{M}}(\bm{x})$. Sensitivity to the number of and choice of $R$'s is considered below.
So, analogous to the GPP, we assume $\bm{T}=(T_{1},\bm{T}_{2})$, where $T_{1}=\log n(\bm{x})$ and the components of $\bm{T}_{2}$ are $T_{2,r}(\bm{x})=\sqrt{\hat{K}_{r}(\bm{x})}$.

\cite{Soubeyrandetal(13)} propose an optimized weight function ABC strategy for functional summary statistics with application to spatial point processes (the Neyman-Scott process and a marked point process). For smaller $M$ ($\leq 20$), they calculate an optimized weighted \emph{distance} by the Nelder-Mead algorithm between the simulated and observed functional statistics by minimizing the mean square error of a point estimate of $\bm{\theta}$.
For larger M, the Nelder-Mead algorithm is not available.  So, \cite{Soubeyrandetal(13)} adopt lower dimensional piecewise constant weight parameters which can be optimized through the algorithm to obtain manageable computation time.

\subsection{Explicit specification of our ABC algorithm}
\label{sec:algorithm}

The ABC algorithm we adopt is based on a semi-automatic approach proposed by \cite{FearnheadPrangle(12)}.
They argue that the optimal choice of $\bm{T}(\bm{y})$ is $E(\bm{\theta}|\bm{y})$ and then discuss how to construct $E(\bm{\theta}|\bm{y})$. They consider a linear regression approach to construct the summary statistics through a pilot run.
In our setting, we generate $L$ sets of $\{\bm{\theta}_{\ell}, \bm{x}_{\ell}\}_{\ell=1}^{L}$ (choice for $L$ discussed below).  Then, we implement a linear regression for $E(\bm{\theta}_{\ell}|\bm{y})=\bm{a}+\bm{b}\bm{\eta}(\bm{x}_{\ell},\bm{y})$ where $\bm{\eta}(\bm{x}_{\ell},\bm{y})$ is a vector of functions of the summary statistics constructed from the simulated and observed point patterns \footnote{\cite{FearnheadPrangle(12)}) implement linear regression for each component of $\bm{\theta}$. Since we have a small number of parameters, we keep the notation as linear regression for multivariate responses.}.
We take $\bm{\eta}(\bm{x}, \bm{y})=(\eta_{1}(\bm{x}, \bm{y}), \bm{\eta}_{2}(\bm{x}, \bm{y}))$ with, for $r=1,...,M$,
\begin{align}
\eta_{1}(\bm{x}, \bm{y})=\log n(\bm{x})-\log n(\bm{y}), \quad \eta_{2,r}(\bm{x}, \bm{y})=\biggl|\sqrt{\hat{K}_{r}(\bm{x})}-\sqrt{\hat{K}_{r}(\bm{y})}\biggl|^2
\end{align}

After obtaining $\hat{\bm{a}}$ and $\hat{\bm{b}}$ by least squares, we can calculate $\hat{\bm{\theta}}_{*}=\hat{\bm{a}}+\hat{\bm{b}}\bm{\eta}(\bm{x}^{*},\bm{y})$ for any simulated $\bm{x}^{*}$.  We set $\hat{\bm{\theta}}_{obs} = \hat{\bm{a}}$ and specify our distance function for the ABC through $\Psi(\hat{\bm{\theta}}^{*},\hat{\bm{\theta}}_{obs})$ with $\Psi$ specified below. To facilitate the regression, we take a log transformation of the parameter vector, e.g., $\bm{\theta}=(\log \beta, \log \gamma)$ for the Strauss process.
Given the results of the pilot run, the approach proposed of \cite{FearnheadPrangle(12)}  implements the ABC-MCMC algorithm by \cite{Marjorametal(03)}. ABC-MCMC is a straightforward extension of the standard MCMC framework to ABC; convergence to the approximate posterior distribution, $\pi_{\epsilon}(\bm{\theta}|\bm{T}(\bm{y}))$, is guaranteed. Specifically, with $t$ denoting iterations and $q(\cdot|\cdot)$ denoting a proposal density,
\begin{itemize}
\item[1.] Let $t=1$.
\item[2.] Generate $\bm{\theta}^{*}\sim q(\bm{\theta}|\bm{\theta}^{(t-1)})$ and $\bm{x}^{*}\sim \pi(\bm{x}|\bm{\theta}^{*})$ and calculate $\hat{\bm{\theta}}^{*}=\hat{\bm{a}}+\hat{\bm{b}}\bm{\eta}(\bm{x}^{*},\bm{y})$. Repeat this step until $\Psi(\hat{\bm{\theta}}^{*},\hat{\bm{\theta}}_{obs})<\epsilon$ where $\hat{\bm{\theta}}_{obs}=\hat{\bm{a}}$ and $\Psi(\hat{\bm{\theta}}^{*},\hat{\bm{\theta}}_{obs})$ is defined below.
\item[3.] Calculate the acceptance ratio
$\alpha=\min\biggl\{1, \frac{\pi(\bm{\theta}^{*})q(\bm{\theta}^{(t-1)}|\bm{\theta}^{*})}{\pi(\bm{\theta}^{(t-1)})q(\bm{\theta}^{*}|\bm{\theta}^{(t-1)})}\biggl\}.$
If $u<\alpha$ where $u\sim \mathcal{U}(0,1)$ retain $\bm{\theta}^{(t)}=\bm{\theta}^{*}$, otherwise $\bm{\theta}^{(t)}=\bm{\theta}^{(t-1)}$. Return to step 2 and $t\to t+1$.
\end{itemize}
As a distance measure, we use the componentwise sum of quadtratic loss for the log of the parameter vector, i.e., $\Psi(\hat{\bm{\theta}}_{\ell}, \hat{\bm{\theta}}_{obs})=\sum_{j}(\hat{\theta}_{\ell,j}-\hat{\theta}_{obs,j})^2/\hat{\text{var}}(\hat{\theta}_{j})$ where $\hat{\text{var}}(\hat{\theta}_{j})$ is the sample variance of $j$-th component of $\hat{\bm{\theta}}$.
To choose an acceptance rate $\epsilon$, through the pilot run we obtain the empirical percentiles of $\{\Psi(\hat{\bm{\theta}}_{\ell},\hat{\bm{\theta}}_{obs})\}_{\ell=1}^{L}$ and then select $\epsilon$ according to these percentiles. Step 2 is the most computationally demanding. We need to simulate the proposed point pattern $\bm{x}^{*}\sim \pi(\bm{x}|\bm{\theta}^{*})$ until $\Psi(\hat{\bm{\theta}}^{*}, \hat{\bm{\theta}}_{obs})<\epsilon$.

With regard to choice of number of summary statistics, when $M$ is large more information is obtained but, if too large, overfitting, relative to the number of points in the point pattern, results.  As a strategy for this selection, we specify $M$ with equally spaced $r$'s, and implement a lasso (\cite{Tibshirani(96)}). We determine the penalty parameter for the lasso by cross-validation, and preserve the regression coefficients corresponding to the optimal penalty by using \texttt{glmnet} (\cite{Friedmanetal(10)}). In our simulation study, we examine sensitivity to the selection of $M$.

We remark that, in frequentist analysis, the minimum contrast estimator is often used.  This is the value of $\bm{\theta}$ which minimizes $\sum_{r} |\hat{K}_{r}(\bm{x})^{a}-K_{\bm{\theta}}(r)^{a}|^{b}$ where $K_{\theta}$ is the theoretical $K$-function, $\hat{K}_{r}(\bm{x})$ is the empirical estimator for the $K$-function, and $a$, $b$ are user-specified parameters (\cite{Diggle(03)}).
The minimum contrast estimator requires analytical forms for the functional statistics which are not necessarily available for repulsive point processes, e.g., the $K$ function for the Mat\'{e}rn-Whittle kernel function requires numerical methods (though the pair correlation function, whose analytical form is available for the Mat\'{e}rn-Whittle kernel function, is an alternative functional summary statistics for the minimum contrast estimator).
ABC does not require analytical expressions for the functional statistics because the approach compares the "estimated" $K$-function for observed and simulated point patterns.
However, if analytical forms for the functional summary statistics are available, the minimum contrast estimator or composite likelihood estimators (\cite{BaddeleyTurner(00)} and \cite{Guan(06)}) would be available and easy to implement. Furthermore, software for these estimators has already been developed (\cite{BaddeleyTurner(05)}).

As a final comment here, we have compared our proposed ABC-MCMC algorithm with the straightforward exchange algorithm of Murray et al. (2006), mentioned above, for a Strauss Gibbs point process.  Without providing details, suppose we compute inefficiency factors (IF) for parameters, i.e., the ratio of the numerical variance of the estimate from the MCMC samples relative to that from hypothetical uncorrelated samples, using both model fitting approaches.  We find that the IF for the exchange algorithm tend to be an order of magnitude greater than those from the ABC-MCMC algorithm.  Also, the ABC-MCMC algorithm allows simple parallelization, not possible for the exchange algorithm, so, computationally, it can be much faster.
\subsection{Model Checking for Repulsive Point Processes}
\label{sec:check}

Model checking for spatial point processes has a relatively small literature.  Approaches include (i) goodness of fit tests that can be used in conjuction with Poisson processes (e.g., \cite{Ripley(88)}) which emerge as special cases of Monte Carlo tests and (ii) residual analysis based on the Papangelou conditional intensity (see \cite{Baddeleyetal(05)}).  For DPPs, likelihood ratio tests are available (see \cite{Lavancieretal(15)}).
However, such tests would not work to compare different types of repulsive point processes, i.e., DPPs with GPPs.
For GPPs, without analytically available likelihoods, simulated comparison of the $K$ or $L$ functions under the fitted model with the observed are offered for model checking.

In the Bayesian framework, a cross validation approach based on training and test datasets  can be applied for point processes with conditionally independent location densities, e.g., Poisson processes and Cox processes (\cite{LeiningerGelfand(16)}). However, this approach is unavailable for repulsive point processes because holding out points will alter the geometric structure, hence the interaction structure, of the point pattern.
As an alternative in our setting, we consider prior predictive Monte Carlo tests using the statistic $s_{r}(\bm{x})$ (\ref{eq:s_R}).
For choices of $r$, we can implement the test for $s_{r}(\bm{y})$ together with the set $\{s_{r}(\bm{x}_{u}^{*}), u=1,\ldots, U\}$ where the $\bm{x}_{u}^{*}$'s are generated under the model with $\bm{\theta}_{u}\sim \pi(\bm{\theta})$.

For model comparison, we consider the ranked probability score (RPS, \cite{GneitingRaftery(07)} and \cite{Czadoetal(09)}) which assesses the performance of a predictive distribution relative to an observation, in our case to an observed count. Intuitively, a good model will provide a predictive distribution that is very concentrated around the observed count.  For a set $B \subset D$,  we calculate the RPS, via Monte Carlo integration, using posterior samples, as
\begin{align}\label{eq:RPS}
RPS(F, N^{test}(B))=\frac{1}{T}\sum_{t=1}^{T}|N^{(t)}(B)-N^{(test)}(B)|-\frac{1}{2T^2}\sum_{t=1}^{T}\sum_{t'=1}^{T}|N^{(t)}(B)-N^{(t')}(B)|.
\end{align}
Applying (\ref{eq:RPS}) to a collection of $B_{j}$ uniformly drawn over $D$ and summing over $j$ gives a model comparison criterion. Smaller values of the sum are preferred. We calculate RPS as in-sample comparison of predictive performance (see \cite{LeiningerGelfand(16)}).
\section{Simulation Studies}
\label{sec:simulation}
We consider simulation to address two points.  In Section \ref{sec:rec}, we provide a proof of concept example to clarify how well we can recover model parameters using our ABC approach.  In Section \ref{sec:assess}, we provide an example to illustrate model assessment.

\subsection{Recovery of Model Parameters}
\label{sec:rec}
\begin{figure}[ht]
  \caption{Simulated realization (left) and profile pseudo likelihood (right) of a Strauss process}
  \label{fig:ex1}
 \begin{minipage}{0.48\hsize}
  \begin{center}
   \includegraphics[width=6.5cm]{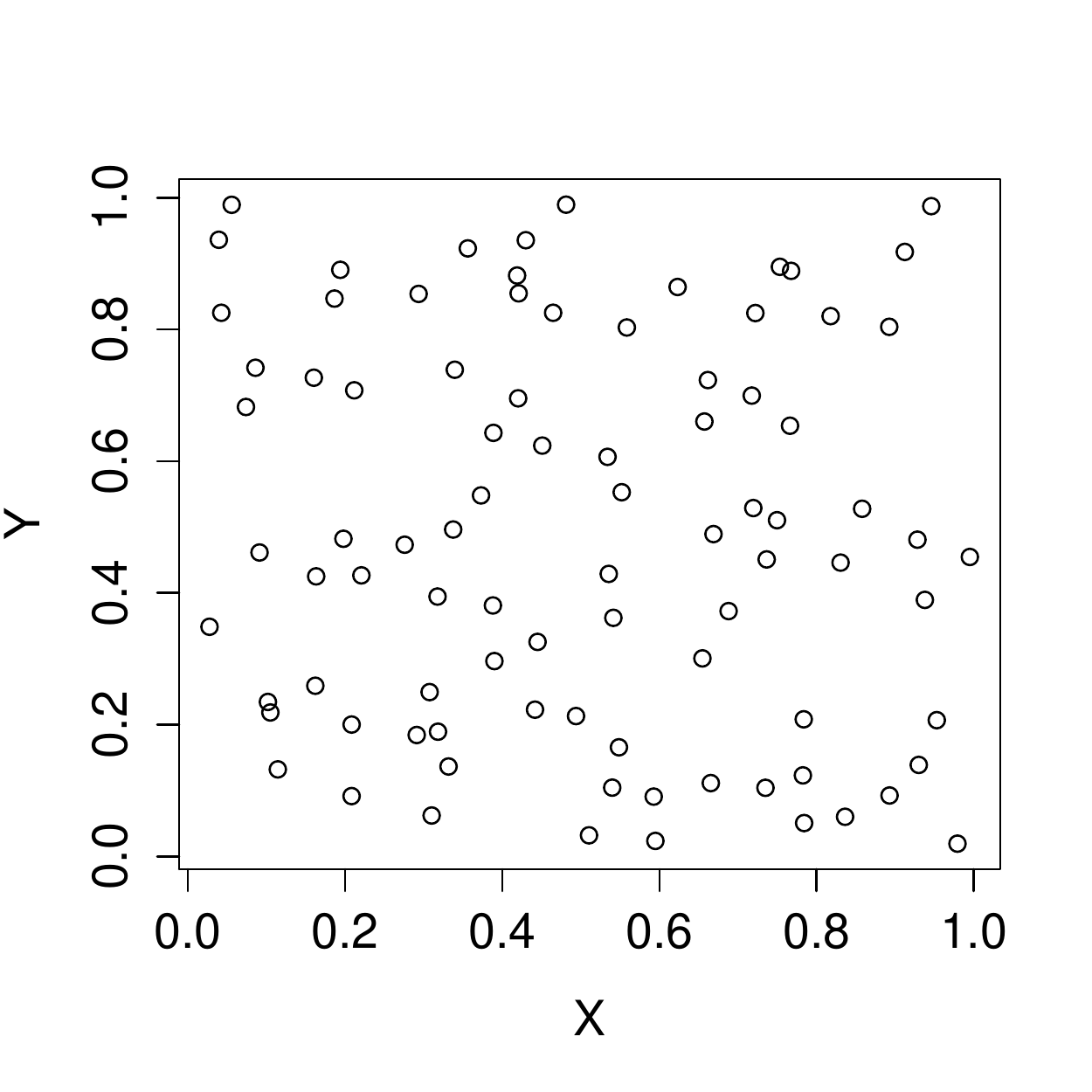}
  \end{center}
 \end{minipage}
 \hfill
 \hfill
 \begin{minipage}{0.48\hsize}
  \begin{center}
   \includegraphics[width=6.5cm]{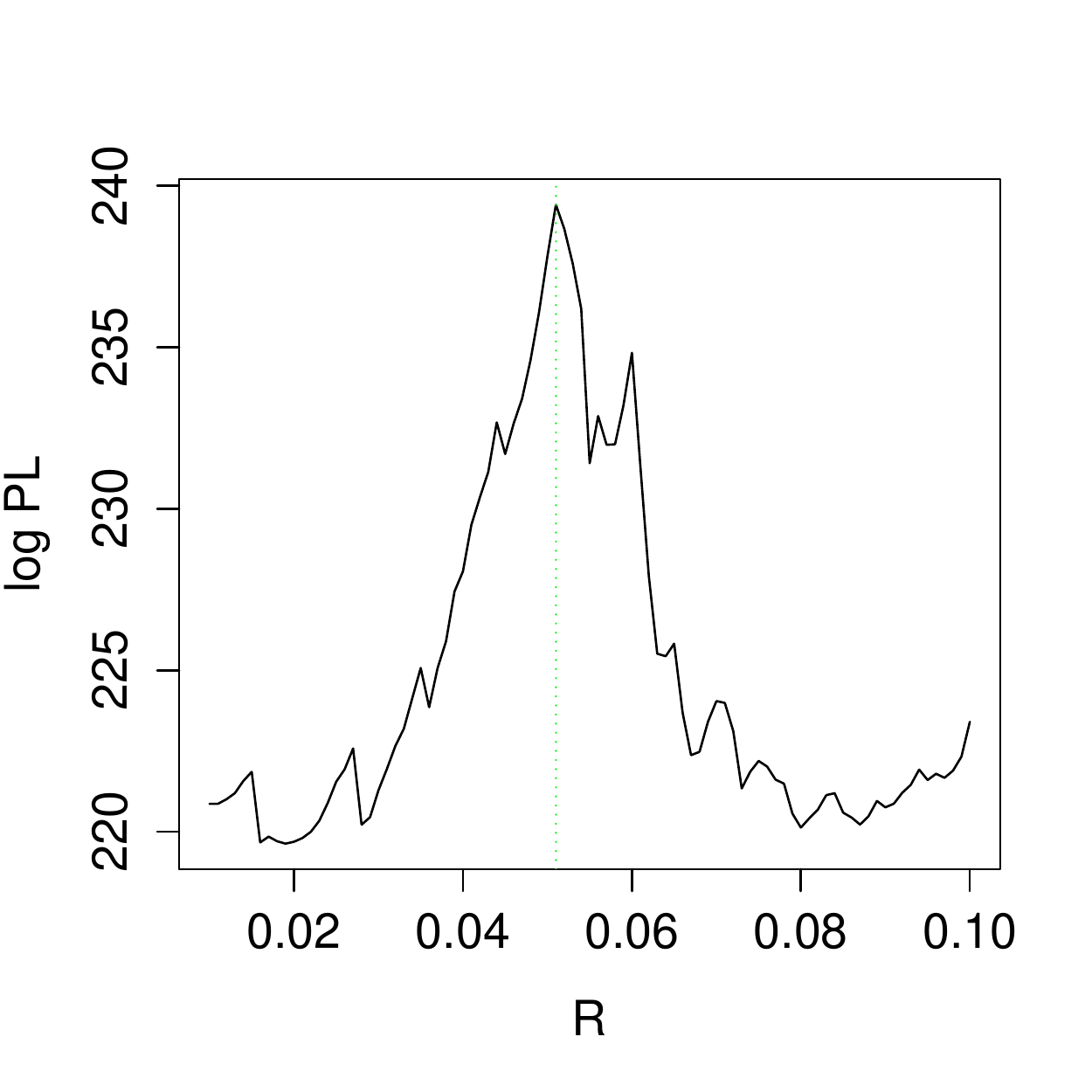}
  \end{center}
\end{minipage}
\end{figure}

First, we consider the strongly repulsive Strauss process on $D=[0,1]\times [0,1]$ as specified in (4),
\begin{equation}
\pi(\bm{x})=\beta^{n(\bm{x})}\gamma^{s_{R}(\bm{x})}/c(\beta,\gamma), \quad s_{R}(\bm{x})=\sum_{\{\xi, \eta\}\subseteq \bm{x}}\bm{1}[\|\xi-\eta\|\le R]
\end{equation}
where $\beta=200$, $\gamma=0.1$, and $R=0.05$.
The realization is shown in the left panel of Figure \ref{fig:ex1}.  The number of points is $n=88$.
The prior specification is a uniform for $\beta$ and $\gamma$: $\pi(\beta)=\mathcal{U}(50, 400)$, $\pi(\gamma)=\mathcal{U}(0, 1)$.
The interaction radius $R$ is estimated by the maximum pseudo-likelihood method. The estimated value is $\hat{R}=0.051$ which is close to the true value (see the right panel of Figure $\ref{fig:ex1}$). We use $\bm{T}_{suff} \equiv (\log n(\bm{x}), \sqrt{K_{\hat{R}}(\bm{x})})$ as our summary statistic.  In the pilot run, we generate $L=$10,000 sets of ($\beta$, $\gamma$, $\bm{x}$) from $\pi(\bm{x}|\beta,\gamma)$ $\times\pi(\beta)\pi(\gamma)$ under $R$ fixed at $\hat{R}$, we calculate regression coefficients $\hat{\bm{a}}$ and $\hat{\bm{b}}$, and we decide the truncation level $\epsilon$ based on the \emph{$p_{*}\%$} estimated percentiles of $\{\Psi(\hat{\bm{\theta}}_{\ell},\hat{\bm{\theta}}_{obs})\}_{\ell=1}^{L}$.
We set $\bar{\hat{\bm{\theta}}}=\hat{\bm{a}}$ as the initial value and preserve 1,000 samples as posterior draws.

\begin{figure}[ht]
  \caption{Simulated realization of a DPP with Gaussian kernel (left) and a DPP with power exponential spectral (right)}
  \label{fig:ex23}
 \begin{minipage}{0.48\hsize}
  \begin{center}
   \includegraphics[width=6.5cm]{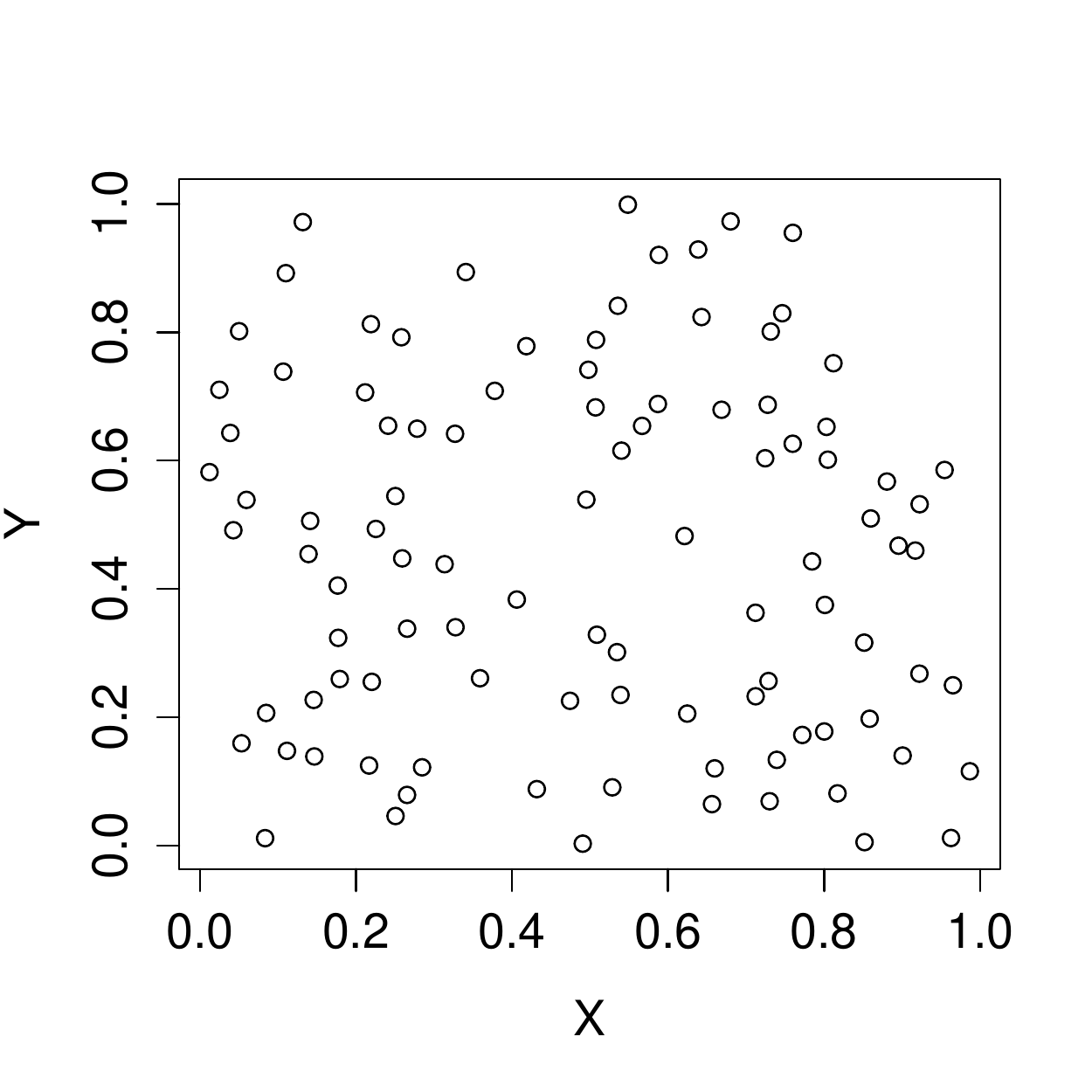}
  \end{center}
\end{minipage}
 \hfill
 \hfill
 \begin{minipage}{0.48\hsize}
  \begin{center}
   \includegraphics[width=6.5cm]{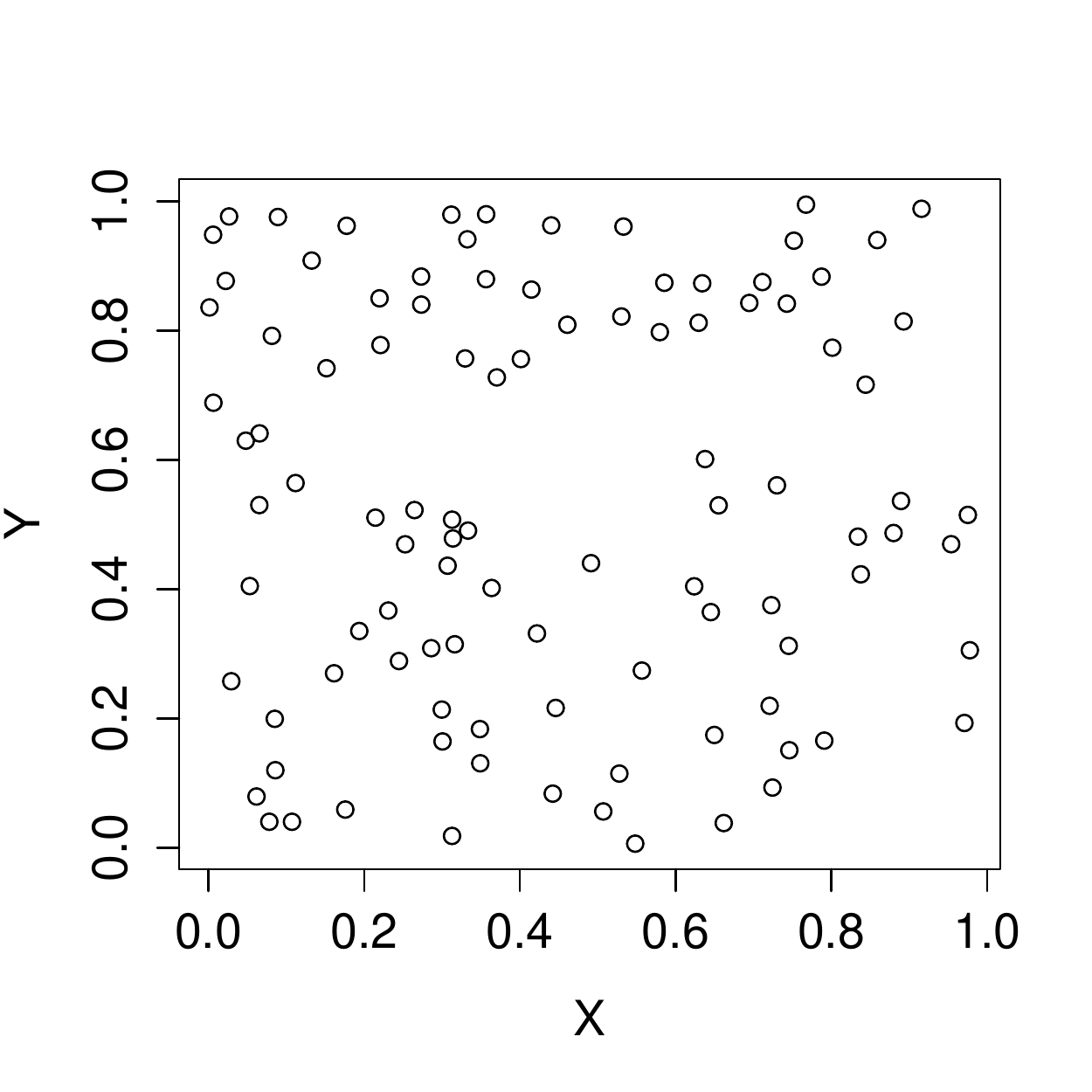}
  \end{center}
\end{minipage}
\end{figure}

Table $\ref{tab:Sim1}$ shows the estimation results. The posterior for $n$ is given though it is not a model parameter.
We show three different acceptance levels. $\bm{T}_{suff}, \epsilon=0$ provides the exact posterior distribution (due to sufficiency, and it can be observed due to discreteness). With $\bm{T}_{suff}, p_{*}=1\%$ and $\bm{T}_{suff}, p_{*}=5\%$, although the variance of estimated parameters increases slightly, the posterior means are well estimated.

\begin{table}[ht]
\caption{Estimation results for the Strauss process}
\label{tab:Sim1}
\centering
\begin{tabular}{lcccccc}
\hline
\hline
Strauss  & True & Mean & Stdev & $95\%$ Int  \\
\hline
 $\bm{T}_{suff}$  $\epsilon=0$ & & & & \\
$\beta$ & 200 & 180.9 & 23.65 & [139.3, 233.5]   \\
$\gamma$ & 0.1 & 0.165 & 0.062 & [0.066, 0.309]  \\
$n$ & 88 & 88 & 0 & [88, 88]  \\
\hline
 $\bm{T}_{suff}$ $p_{*}=1\%$ & & & & \\
$\beta$ & 200 & 180.7 & 24.53 & [135.0, 230.6]   \\
$\gamma$ & 0.1 & 0.167 & 0.071 & [0.057, 0.336]  \\
$n$ & 88 & 88 & 1.593 & [86, 91]   \\
\hline
 $\bm{T}_{suff}$ $p_{*}=5\%$ & & & & \\
$\beta$ & 200 & 181.5 & 31.08 & [127.1, 247.0]  \\
$\gamma$ & 0.1 & 0.157 & 0.091 & [0.020, 0.361]   \\
$n$ & 88 & 88 & 5.705 & [77, 97]  \\
\hline
\hline
\end{tabular}
\end{table}

As second example, we consider a DPP with a Gaussian covariance kernel (DPP-G),
\begin{align}
C_{G}(x,y)=\tau\exp\biggl(-\frac{\|x-y\|^2}{\sigma^2} \biggl), \quad \tau=100, \quad \sigma=0.05.
\end{align}
The realization is shown in the left panel of Figure \ref{fig:ex23} with the random $n=100$.
A uniform distribution for both parameters is assumed: $\pi(\tau)=\mathcal{U}(50, 200)$ (an interval which includes $\hat{\tau} = n/|D|=100$) and $\pi(\sigma)=\mathcal{U}(0.001, \sigma_{max})$, where $\sigma_{max}=1/\sqrt{\pi\tau}\approx0.0564$ (required  for the existence of a DPP with a Gaussian kernel).
For DPP models, $\bm{T}_{func}(\bm{x})=(\log n(\bm{x}), \sqrt{K_{r_{1}}(\bm{x})},\ldots, \sqrt{K_{r_{M}}(\bm{x})})$.
We consider $M=20$ and $10$ equally spaced values over $[0, 0.1]$, i.e., $(r_{1}=0.005, r_{2}=0.01,\ldots, r_{20}=0.1)$ and $(r_{1}=0.01, r_{2}=0.02,\ldots,r_{10}=0.1)$.
We implement a pilot run and then our ABC algorithm with the same number of posterior samples as in the first example. Table \ref{tab:Sim23} illustrates that the true values of parameters are recovered well.
The estimation results are insensitive to the choice of $M$.
The posterior variance of the parameters increases slightly when the tolerable level based on $p_{*}$ increases but the posterior means are well estimated.

As third example, we consider a DPP with power exponential spectral density (DPP-PE),
\begin{align}
\varphi(x)=\tau \frac{\alpha^{2}}{\pi \Gamma(2/\nu+1)}\exp(-\|\alpha x\|^{\nu}), \quad \tau=100, \quad \alpha=0.1, \quad \nu=10 \text{(fixed)}
\end{align}
This kernel can capture stronger repulsiveness than a DPP with a Gaussian kernel (DPP-G). Though the parameters in this kernel are not interpretable, we can still investigate whether or not they are recovered. The random number of points is $n=101$. A uniform distribution for $\tau$ and $\alpha$ is assumed: $\pi(\tau)=\mathcal{U}(50, 200)$ and $\pi(\alpha)=\mathcal{U}(0.001, \alpha_{max})$ where $\alpha_{max}=\sqrt{\Gamma(2/\nu+1)\pi/\tau}\approx 0.1698$.
Again, we implement a pilot run and then our ABC algorithm with the same number of posterior samples. Table \ref{tab:Sim23} demonstrates that the true values of parameters are recovered well.

\begin{table}[ht]
\caption{Estimation results for the DPP with Gaussian kernel (left) and with power exponential spectral density (right) for each number of evaluation grids and percentiles.}
\label{tab:Sim23}
\centering
\begin{tabular}{lccccccccccc}
\hline
\hline
Gauss  & True & Mean & Stdev & $95\%$ Int & PE & True & Mean & Stdev & 95$\%$ Int \\
$(M, p_{*}\%)$  & & & & &&&&&\\
\hline
  $(10, 1\%)$  & & & & &&&&&\\
$\tau$ & 100 & 100.4 & 5.863 & [89.84, 112.5]  & $\tau$ & 100 & 100.9 & 6.011 & [88.47, 112.6] \\
$\sigma$ & 0.05 & 0.047 & 0.006 & [0.030, 0.057] & $\alpha$ & 0.1& 0.113 & 0.022 & [0.069, 0.155]  \\
$n$ & 100 & 100 & 3.198 & [94, 106] & $n$ & 101 & 101 & 2.99 & [96, 107]  \\
\hline
  $(10, 5\%)$ & & & & &&&&& \\
$\tau$ & 100 & 100.0 & 8.205 & [85.26, 117.1] & $\tau$ & 100 & 102.9 & 7.899 & [87.93, 117.3]  \\
$\sigma$ & 0.05 & 0.045 & 0.008 & [0.026, 0.057] & $\alpha$ & 0.1 & 0.112 & 0.025 & [0.060, 0.160]  \\
$n$ & 100 & 100 & 7.026 & [88, 113] & $n$ & 101 & 102 & 6.307 & [90, 113] \\
\hline
  $(20, 1\%)$ & & & & &&&&&\\
$\tau$ & 100 & 99.28 & 6.277 & [86.30, 110.7] & $\tau$ & 100 & 102.3 & 6.568 & [88.93, 114.8]  \\
$\sigma$ & 0.05 & 0.048 & 0.006 & [0.032, 0.057] & $\alpha$ & 0.1 & 0.110 & 0.22 & [0.061, 0.151]  \\
$n$ & 100 & 100 & 3.140 & [94, 105] & $n$ & 101 & 101 & 2.922 & [96, 106]  \\
\hline
  $(20, 5\%)$ & & & & &&&&& \\
$\tau$ & 100 & 99.06 & 7.434 & [84.32, 113.6] & $\tau$ & 100 & 101.4 & 7.490 & [86.02, 116.0]\\
$\sigma$ & 0.05 & 0.045 & 0.007 & [0.028, 0.057] & $\alpha$ & 0.1 & 0.110 & 0.025 & [0.057, 0.157] \\
$n$ & 100 & 99 & 6.596 & [88, 111] & $n$ & 101 & 101 & 6.313 & [90, 113] \\
\hline
\hline
\end{tabular}
\end{table}

\subsection{Model Assessment}
\label{sec:assess}

As in the discussion in Section \ref{sec:check}, we examine model adequacy through Monte Carlo tests.
We consider two different point pattern sizes: (A) $n\approx 100$ and (B) $n\approx 500$.
Four true datasets are generated for each pattern size.  For (A): (A.i) HPP $(\lambda=100)$, (A.ii) Strauss process $(\beta=250, \gamma=0.05, R=0.05)$, (A.iii)  DPP-G $(\tau=100, \sigma=\sigma_{max})$, (A.iv) DPP-PE $(\tau=100, \alpha=\alpha_{max}, \nu=10)$. For (B): (B.i) HPP $(\lambda=500)$, (B.ii) Strauss process $(\beta=1000, \gamma=0.05, R=0.02)$, (B.iii) DPP-G $(\tau=500, \sigma=\sigma_{max})$ and (B.iv) DPP-PE $(\tau=500, \alpha=\alpha_{max}, \nu=10)$. The DPP-PE with $\nu=10$ and $\alpha=\alpha_{max}$ provides stronger repulsion than the DPP-G model with $\sigma=\sigma_{max}$ (see, \cite{Lavancieretal(15)}).
All settings are expected to have 100 points for (A) and 500 points for (B), respectively so we focus on comparing the second order statistics $s_{r}(\bm{x})$.

For each model, the prior specifications are as follows.
Let $\mathcal{G}(,)$, $\mathcal{B}(,)$ and $\mathcal{U}(,)$ denote Gamma, Beta and uniform distributions respectively.  Then, for (A), HPP: $\pi(\lambda)=\mathcal{G}(200,2)$  $(E[\lambda]=100, \quad \text{Var}[\lambda]=50)$, Strauss: $\pi(\beta)=\mathcal{U}(75,400)$ $\pi(\gamma)=\mathcal{B}(1,6)$, DPP-G: $\pi(\tau)=\mathcal{G}(200,2)$ $\pi(\sigma/\sigma_{max})=\mathcal{B}(6,1)$, DPP-PE: $\pi(\tau)=\mathcal{G}(200,2)$ $\pi(\alpha/\alpha_{max})=\mathcal{B}(6,1)$.
The beta priors for $\gamma$, $\alpha/\alpha_{max}$ and $\sigma/\sigma_{max}$ imply moderate to strong repulsion within each model because each model shows stronger repulsiveness when $\gamma$ is small and the $\alpha/\alpha_{max}$ and $\sigma/\sigma_{max}$ are large.

For (B) HPP: $\pi(\lambda)=\mathcal{G}(1000,2)$  $(E[\lambda]=500, \quad \text{Var}[\lambda]=250)$, Strauss: $\pi(\beta)=\mathcal{U}(350,1200)$ $\pi(\gamma)=\mathcal{B}(1,6)$, DPP-G: $\pi(\tau)=\mathcal{G}(1000,2)$ $\pi(\sigma/\sigma_{max})=\mathcal{B}(6,1)$ where $\sigma_{max}=1/\sqrt{\pi \tau}$, DPP-PE: $\pi(\tau)=\mathcal{G}(1000,2)$ $\pi(\alpha/\alpha_{max})=\mathcal{B}(6,1)$.

Table $\ref{tab:ma1}$ shows estimated $p$-values for the Monte Carlo tests associated with the observed $s_{r}(\bm{y})$ by simulating 999 point patterns $\{\bm{y}^{*}\}$ from the prior predictive distribution of each model. For $n \approx 100$, models are assessed at the radii $r=(0.03, 0.05, 0.07, 0.09)$;  for $n \approx 500$ models are assessed at the radii $r=(0.01, 0.02, 0.03, 0.04)$.
For $n\approx 100$ and $n\approx 500$ we see similar results.
When the true model is the HPP or the Strauss process, the other models are sometimes formally criticized but, regardless, show smaller $p$-values. When the DPP-G is the true model, only the HPP is criticized. When DPP-PE is true, DPP-G and HPP are criticized but Strauss is not. This result is not surprising  because, though the DPP-PE presents stronger repulsiveness than the DPP-G, the Strauss process shows very strong repulsiveness. As expected, the results emerge more strongly for the larger point patterns.

\begin{figure}[htbp]
\caption{In sample relative RPS: $n\approx 100$ (left) and $n\approx 500$ (right)}
  \begin{center}
   \includegraphics[width=15cm]{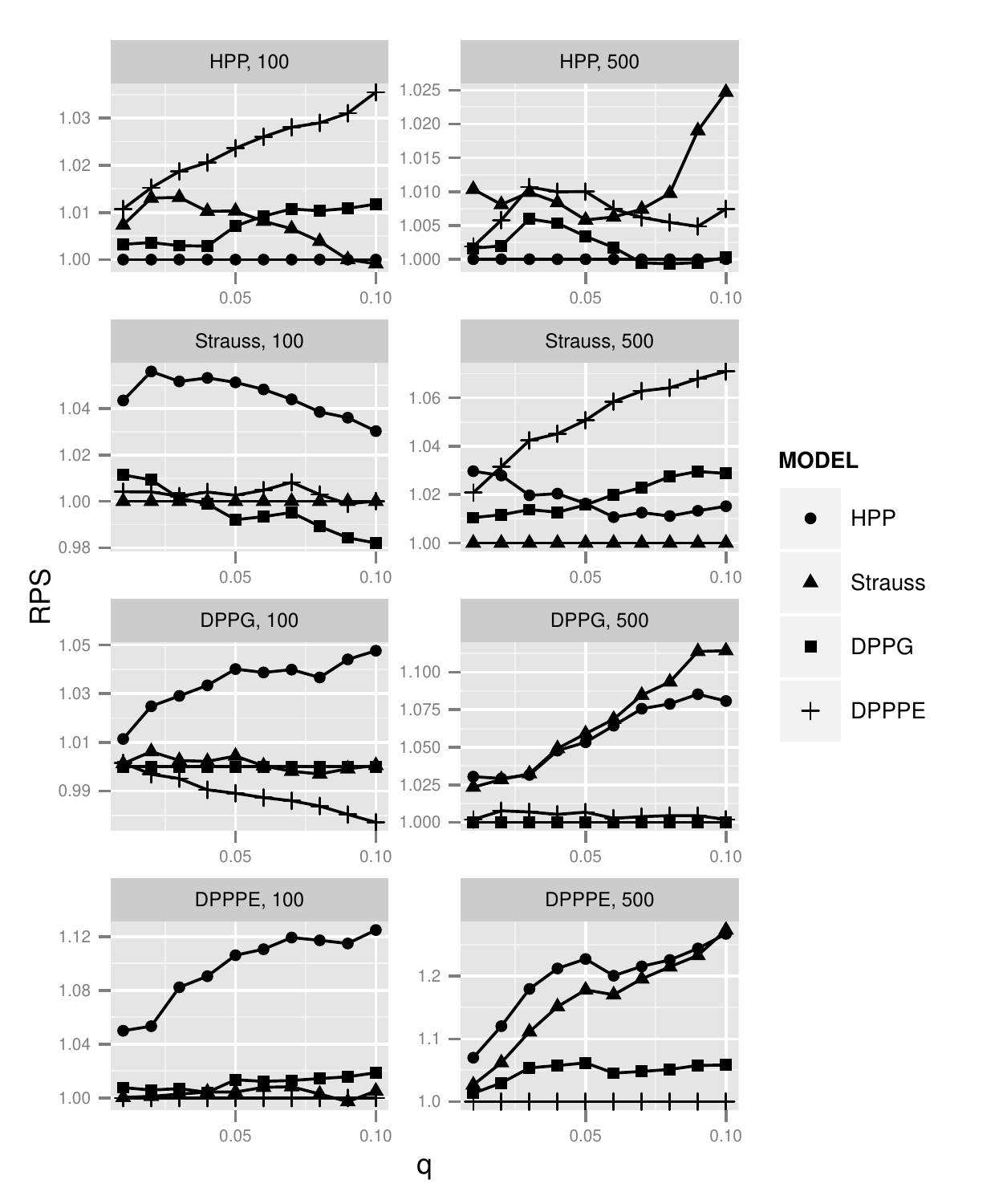}
  \end{center}
  \label{fig:rps_sim}
\end{figure}

We also investigate in-sample RPS to compare the model fitting.
We set $J=1000$ and sample squares $B_{j}$ uniformly over $D$.  Each $B_{j}$ is a square of size $q|D|$ with $q\in (0, 0.1)$. We calculate RPS for each $B_{j}$ and average over $j=1,\ldots, J$.
Figure $\ref{fig:rps_sim}$ shows the relative RPS, i.e., the ratio of RPS for a particular model relative to the RPS for the true model. For $n \approx 500$, though the ratios are close to $1.00$, the true model is generally preferred. However, when the DPP-G is true, the difference between DPP-G and DPP-PE is small. For $n\approx 100$, the comparison is a bit less successful.  When the HPP or the DPP-PE are true, these models are preferred by the in-sample RPS. However, when the Strauss or the DPP-G are true, they don't show the smallest RPS, although the difference is small.  Moreover, even for $n\approx 100$, the HPP can be distinguished from other repulsive point processes even with moderate repulsiveness.  Altogether, we conclude that a larger number of points will be required to distinguish the true repulsive point process from the other repulsive point processes.

\begin{table}[htbp]
\caption{Model Adequacy: estimated $p$-values for simulated datasets, $n\approx 100$ (left) and $n\approx 500$ (right). Bold denotes significance at $p \leq .1$.}
\label{tab:ma1}
\centering
\begin{tabular}{lccccccccccc}
\hline
\hline
$n\approx 100\backslash r$  & $0.03$ & $0.05$ & $0.07$ & $0.09$ & $n\approx 500\backslash r$ & $0.01$ & $0.02$ & $0.03$ & $0.04$  \\
\hline
HPP  &  &  &  &  & HPP &  &  &  &  \\
\hline
 HPP   & 0.669 & 0.588 & 0.592 & 0.578 &  HPP  & 0.404 & 0.638 & 0.596 & 0.636 \\
Strauss  & \textbf{0.012} & \textbf{0.011} & \textbf{0.077} & 0.137 & Strauss  & \textbf{0.017} & \textbf{0.009} & \textbf{0.086} & 0.147 \\
DPP-G  & \textbf{0.005} & \textbf{0.035} & 0.130 & 0.215 & DPP-G  & \textbf{0.001} & \textbf{0.001} & \textbf{0.013} & \textbf{0.028} \\
DPP-PE  & \textbf{0.067} & 0.149 & 0.217 & 0.257 & DPP-PE  & \textbf{0.001} & \textbf{0.001} & \textbf{0.001} & \textbf{0.034} \\
\hline
\hline
Strauss  &  &  &  &  & Strauss &  &  &  &  \\
\hline
 HPP & \textbf{0.001} & \textbf{0.001} & \textbf{0.092} & 0.268 & HPP & \textbf{0.001} & \textbf{0.001} & \textbf{0.001} & \textbf{0.038} \\
Strauss & 0.481 & 0.450 & 0.697 & 0.719 & Strauss & 0.409 & 0.382 & 0.617 & 0.638 \\
DPP-G & 0.105 & \textbf{0.001} & 0.345 & 0.469 & DPP-G & 0.114 & \textbf{0.001} & 0.239 & 0.369 \\
DPP-PE & 0.127 & \textbf{0.001} & 0.333 & 0.452  & DPP-PE  & 0.413 & \textbf{0.001} & 0.745 & 0.735 \\
\hline
\hline
DPP-G  &  &  &  &  & DPP-G &  &  &  &  \\
\hline
HPP & \textbf{0.005} & \textbf{0.020} & 0.108 & 0.236 & HPP & \textbf{0.001} & \textbf{0.001} & \textbf{0.001} & \textbf{0.034} \\
Strauss & 0.386 & 0.101 & 0.285 & 0.303 & Strauss & 0.409 & 0.768 & 0.380 & 0.492 \\
DPP-G & 0.245 & 0.274 & 0.364 & 0.441 & DPP-G & 0.328 & 0.199 & 0.260 & 0.355 \\
DPP-PE  & 0.224 & 0.275 & 0.357 & 0.421 & DPP-PE  & 0.709 & 0.788 & 0.764 & 0.721 \\
\hline
\hline
DPP-PE  &  &  &  &  & DPP-PE &  &  &  &
 \\
\hline
HPP & \textbf{0.001} & \textbf{0.003} & \textbf{0.017} & \textbf{0.090} & HPP & \textbf{0.001} & \textbf{0.001} & \textbf{0.001} & \textbf{0.001} \\
Strauss & 0.481 & 0.833 & 0.502 & 0.547 & Strauss & 0.409 & 0.768 & 0.380 & 0.492 \\
DPP-G & 0.105 & \textbf{0.075} & 0.104 & 0.201 & DPP-G & 0.114 & \textbf{0.005} & \textbf{0.003} & \textbf{0.036} \\
DPP-PE  & 0.127 & 0.122 & 0.131 & 0.209 & DPP-PE  & 0.413 & 0.160 & 0.144 & 0.199 \\
\hline
\hline
\end{tabular}
\end{table}

\section{Real Data Application}
\label{sec:real}

\begin{figure}[htbp]
\caption{Plot of Duke Forest dataset (left),  profile pseudolikelihood for the Strauss process (middle) and $\hat{L}(r)-r$ for the focused region (right)}
  \label{fig:bw40}
 \begin{minipage}{0.32\hsize}
  \begin{center}
   \includegraphics[width=5cm]{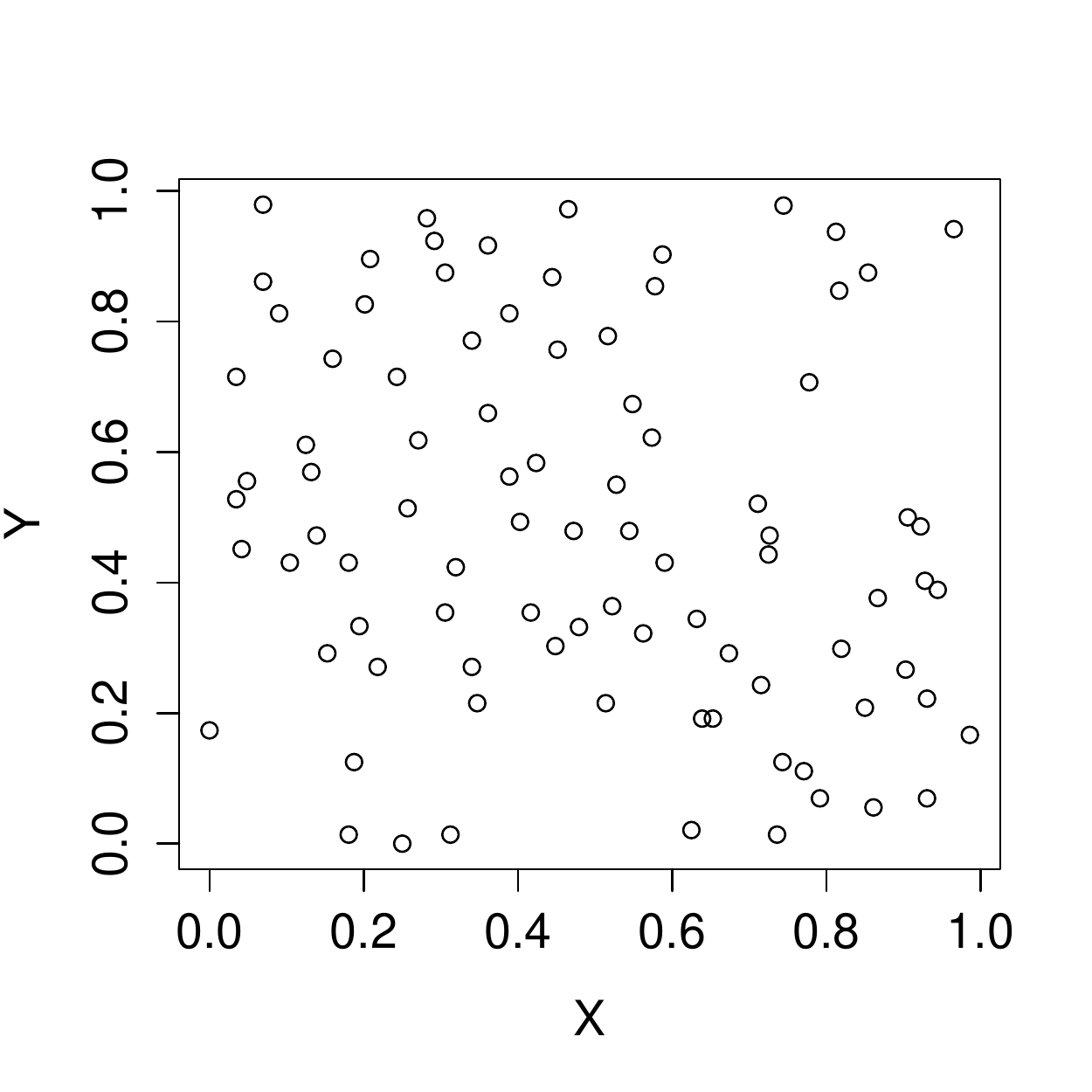}
  \end{center}
 \end{minipage}
 \hfill
 \hfill
 \begin{minipage}{0.32\hsize}
  \begin{center}
   \includegraphics[width=5cm]{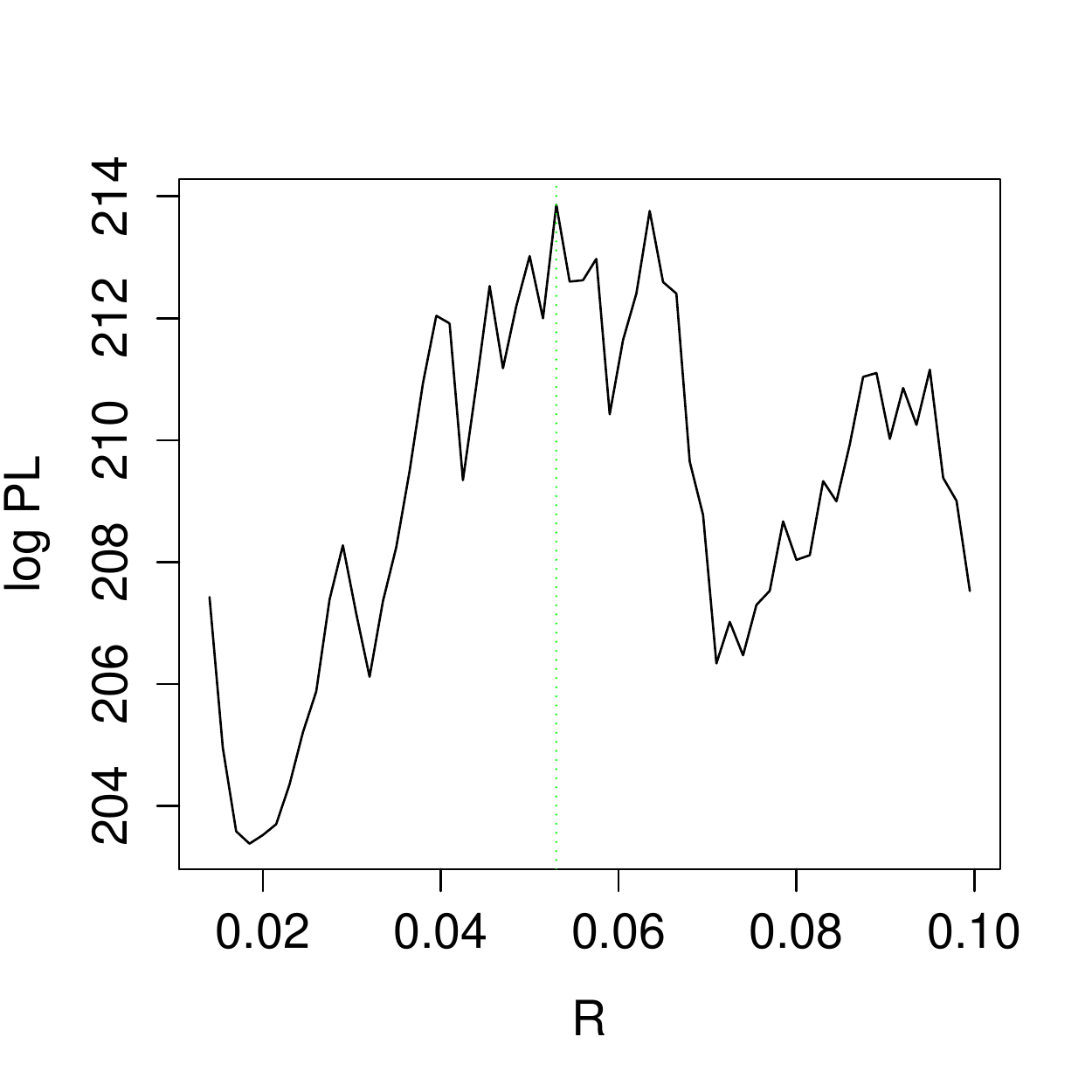}
  \end{center}
\end{minipage}
 \hfill
 \hfill
 \begin{minipage}{0.32\hsize}
  \begin{center}
   \includegraphics[width=5cm]{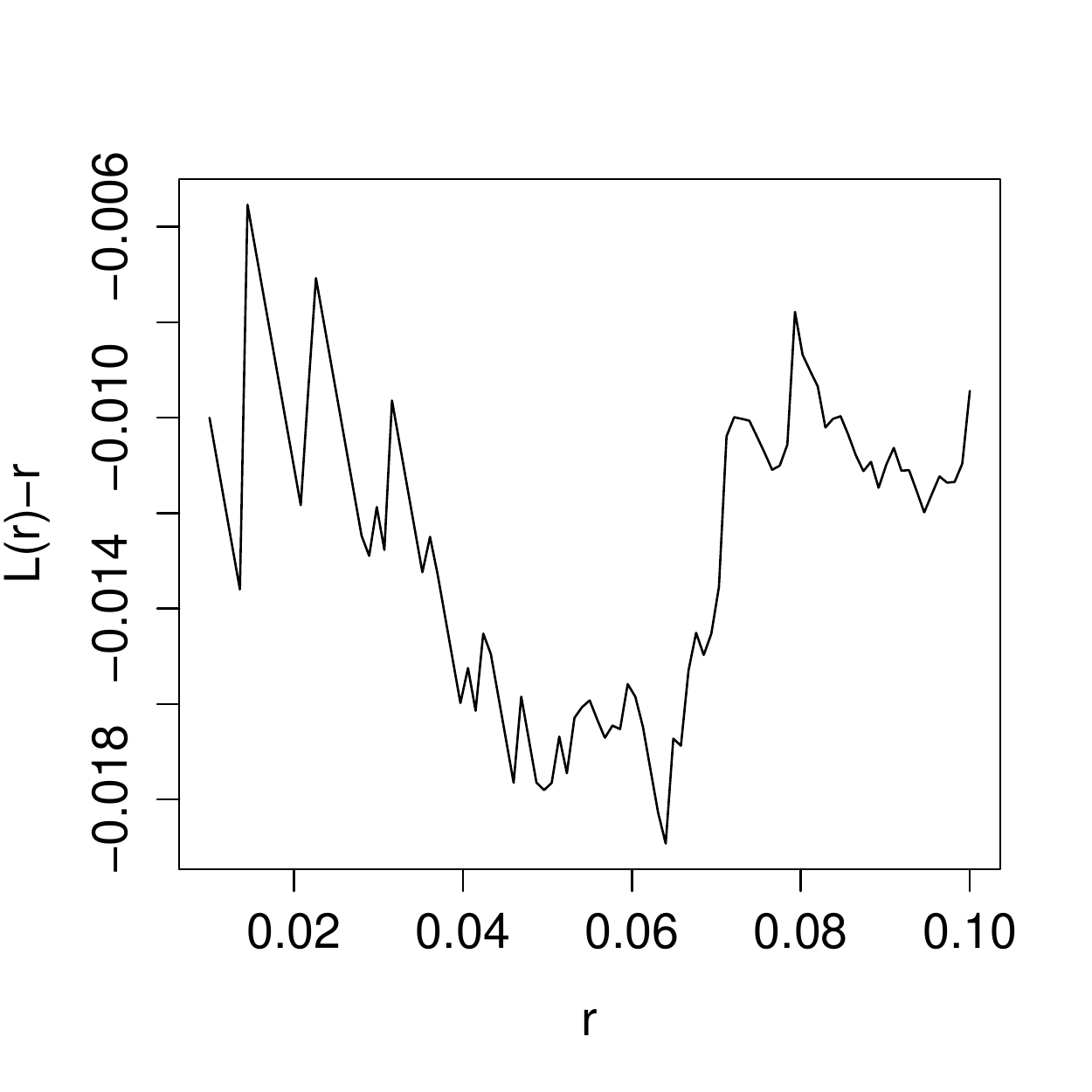}
  \end{center}
\end{minipage}
\end{figure}

We implement our approach for a dataset of tree locations in Duke Forest, comprising 68 species, a total of 13,655 individuals with diameter at breast height (dbh) and location recorded for each. We aggregate the species and remove trees under 40 dbh from the dataset because only older trees exhibit inhibition/repulsion.
The left panel in Figure $\ref{fig:bw40}$ shows the locations of trees whose dbh is larger than 40 over a selected rectangle (rescaled to the unit square for fitting) in the Blackwood region of Duke Forest. The resulting number of points is 89.

For the Strauss model, we include both an interaction radius $R\approx0.053$ and a hardcore radius which are chosen from profile pseudo likelihoods (see, the middle panel in Figure $\ref{fig:bw40}$)\footnote{In the simulation examples, we considered only an interaction radius. However, with real data, often we find hardcore repulsion with a very small radius or moderate repulsion with a larger radius.}.
We also investigated the cases $R=0.02$ and $R=0.035$.
The prior specifications for the Strauss process are: $\beta\sim \mathcal{U}(50,350)$, $\gamma\sim \mathcal{U}(0,1)$. We fix $p_{*}=1\%$ percentiles to determine $\epsilon$ for Strauss models.
We also consider a DPP-G and a DPP-PE with $\tau \sim \mathcal{U}(50, 200)$ and $\sigma\sim \mathcal{U}(0.001,\sigma_{max})$ where $\sigma_{max}=1/\sqrt{\pi\tau}$ for the DPP-G and $\tau \sim \mathcal{U}(50, 200)$ and $\alpha\sim \mathcal{U}(0.001,\alpha_{max})$ where $\alpha_{max}=\sqrt{\Gamma(2/\nu+1)\pi/\tau}$ for the DPP-PE.
Through the pilot run, we fix $M=10$ and $p_{*}=1\%$ percentiles to determine $\epsilon$ for DPP models.  We preserve 1,000 samples for each model.
Table $\ref{tab:real}$ shows the estimation results for the HPP, the Strauss, the DPP-G and the DPP-PE models. The estimated value of $\gamma$ for the Strauss process with $R=0.053$ reveals a moderate level of repulsion.

\begin{table}[htbp]
\caption{Estimation results for the real data}
\label{tab:real}
\centering
\begin{tabular}{lccccccccc}
\hline
\hline
   & Mean & Stdev & $95\%$ Int & & Mean & Stdev & $95\%$ Int \\
\hline
 HPP & & &  & Strauss & & & \\
  & & &  & $R=0.02$ & & & \\
$\beta$  & 89.75 & 5.464 & [79.73, 100.4]  & $\beta$  & 96.55 & 10.34 & [77.11, 118.3] \\
$n$  & 90 & 10.79 & [70, 112] & $\gamma$  & 0.452 & 0.243 & [0.065, 0.937]   \\
  &  &  & & $n$  & 89 & 3.079 & [84, 94]\\
\hline
 DPP-G & & & & Strauss &&& \\
  & & & & $R=0.035$ &&& \\
$\tau$  & 88.96 & 6.446 & [76.94, 101.8] & $\beta$  & 111.6 & 13.25 & [86.99, 138.5]  \\
$\sigma$  & 0.048 & 0.009 & [0.025, 0.060] & $\gamma$  & 0.467 & 0.185 & [0.169, 0.866]  \\
$n$  & 89 & 4.287 & [82, 97]  & $n$  & 90 & 2.446 & [85, 93] \\
\hline
 DPP-PE & & & & Strauss &&& \\
  & & & & $R=0.053$ &&& \\
$\tau$  & 89.47 & 4.936 & [79.16, 98.71]  & $\beta$  & 149.0 & 23.43 & [110.3, 200.3] \\
$\alpha$  & 0.144 & 0.025 & [0.084, 0.179] & $\gamma$  & 0.452 & 0.140 & [0.202, 0.744]  \\
$n$  & 90 & 3.058 & [84, 95]  & $n$  & 90 & 1.109 & [88, 92] \\
\hline
\hline
\end{tabular}
\end{table}


\begin{table}[htbp]
\caption{Model Adequacy: estimated $p$-values for the real dataset. Bold denotes significance at $p \leq .1$}
\label{tab:ma3}
\centering
\begin{tabular}{lcccccc}
\hline
\hline
 Model$\backslash r$  & $0.03$ & $0.05$ & $0.07$ & $0.09$ \\
\hline
 HPP  & \textbf{0.0428} & \textbf{0.0130} & 0.1002 & 0.1662 \\
Strauss $(R=0.02)$ & \textbf{0.0572} & \textbf{0.0528} & \textbf{0.0919} & 0.1040 \\
Strauss $(R=0.035)$ & 0.2373 & \textbf{0.0978} & 0.1292 & 0.1392 \\
Strauss $(R=0.053)$ & 0.3574 & 0.3923 & 0.3340 & 0.3042 \\
DPP-G  & 0.1894 & 0.1311 & 0.1978 & 0.2097 \\
DPP-PE  & 0.2828 & 0.1801 & 0.2274 & 0.2305 \\
\hline
\hline
\end{tabular}
\end{table}

\begin{figure}[htbp]
\caption{In-sample relative RPS for Duke forest data}
  \begin{center}
   \includegraphics[width=12cm]{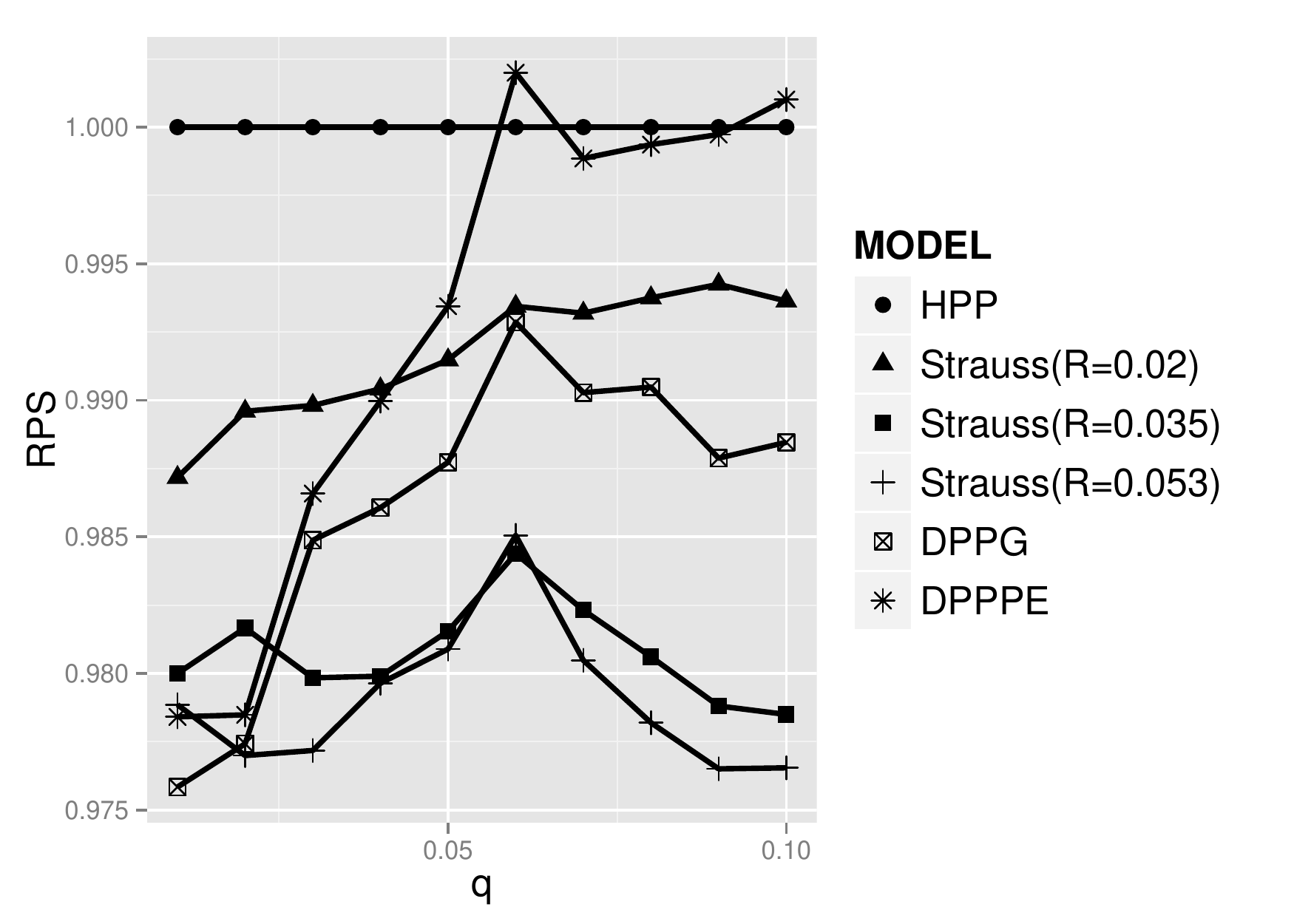}
  \end{center}
  \label{fig:rps_real}
\end{figure}

As above, we consider prior predictive checks using Monte Carlo tests employing $s_{r}(\bm{y})$ for comparing second order properties.
Table $\ref{tab:ma3}$ presents the estimated $p$-values for the observed $s_{r}(\bm{y})$ for each model and radius.
Including prior specification, Table $\ref{tab:ma3}$ reveals that the HPP is criticized for small radii. The Strauss model with small radii ($R=0.02$ and $R=0.035$) are also criticized.
There is evidence in support of repulsion with moderate radius.
We also calculate in-sample RPS.
Figure $\ref{fig:rps_real}$ shows the relative RPS to the RPS of the HPP. Again, the ratios are near $1.00$.  However, despite the somewhat small size of the point pattern, the figure indicates preference for the repulsive point processes, most strongly for the Strauss model with $R=0.053$.

\section{Summary and future work}
\label{sec:disc}

Bayesian inference for DPPs and GPPs is not unified because of challenging model fitting problems specific to each type of process model.
Here, we have proposed a unifying approach for model fitting using ABC. It is attractive because simulation from the models is easier than evaluation of the exact likelihood and because informative summary statistics (first and second moments) are available.
We also offered model assessment strategies for repulsive point processes using Monte Carlo tests and in-sample RPS.
Simulation studies illustrate that true models can be recovered but that it may be difficult to criticize the true repulsive point process in favor of other repulsive specifications when the number of points is small.

Future work will examine nonhomogeneous repulsive point processes.
Because the second order functional summary statistics are available, e.g., the nonhomogeneous versions of the $K$ function (\cite{Baddeleyetal(00)}), our ABC approach can be directly extended. Another challenge is to consider very large datasets. The main computational cost of our algorithm is simulation of $\bm{x}$ for each iteration of the ABC-MCMC.  One promising solution is to run multiple MCMC chains in parallel  (\cite{Wilkinson(05)}).
Through the pilot run, we can obtain the rough approximate posterior mean of the parameters $\bar{\hat{\bm{\theta}}}=\hat{\bm{a}}$. Starting the algorithm with this initial value results in no need for a burn-in period.
In this regard, other types of ABC algorithms can be considered, for example, sequential versions of ABC (ABC-SMC) \cite{Beaumontetal(09)}, \cite{Tonietal(09)} and \cite{Marinetal(12)} which might be more suitable for parallelization.

\section*{Acknowledgements}
The work of the first author was supported in part by the Nakajima Foundation. The authors thank James Clark for providing the Duke Forest dataset.

\bigskip
\begin{center}
{\large\bf SUPPLEMENTARY MATERIAL}
\end{center}

\begin{description}
\item[R-code] R-code is attached for Section $\ref{sec:simulation}$ and $\ref{sec:real}$
\item[Duke Forest data set:] Data set used in Section $\ref{sec:real}$
\end{description}

\bibliographystyle{chicago}
\bibliography{SP}

\end{document}